\documentclass[draft]{agujournal2019}
\usepackage{url} %this package should fix any errors with URLs in refs.
\usepackage{lineno}
\usepackage{soul}
\usepackage{comment}
\usepackage[normalem]{ulem}
\draftfalse
\usepackage{graphicx}
\usepackage{epstopdf}
\usepackage{tikz}
\usepackage[labelformat=simple,position=top]{subfig}
\usepackage{units}
\usepackage{amsmath}   
\usepackage{amssymb}% Align environment...
\usepackage{float}                                         % Provides H float modifier option.
\usepackage{footnote}
\usepackage{lineno}
\usetikzlibrary{shapes.misc}
\tikzset{cross/.style={cross out, draw=black}}
\usepackage{color}                                         
\usepackage{bm}                                            % Bold mathematics symbols. Defines command \bm.
\usepackage{amsbsy}                                        % Bold mathematics symbols.
\newcommand{\mitbf}[1]{\hbox{\mathversion{bold}$#1$}}
\newcommand{\mo}{\mu_0}

\newcommand{\curl}{\nabla \times}
\newcommand{\grad}{\mathrm{grad\,}}

\newcommand{\vz}{\mitbf{0}}
\newcommand{\B}{\mitbf{B}}
\newcommand{\Bc}{\mitbf{B}_\mathrm{C}}
\newcommand{\Boj}{\widetilde{\B}}
\newcommand{\Bjs}{\widehat{\B}}
\newcommand{\Btot}{\bar{\B}}
\newcommand{\BT}{\B_\mathrm{T}}
\newcommand{\BojT}{\Boj_\mathrm{T}}
\newcommand{\BjsT}{\Bjs_\mathrm{T}}
\newcommand{\BoT}{\Bo_\mathrm{T}}
\newcommand{\Bojext}{\Boj^{\mathrm{(e)}}}
\newcommand{\Bjsext}{\Bjs^{\mathrm{(e)}}}
\newcommand{\Boext}{\Bo^{\mathrm{(e)}}}
\newcommand{\Bext}{\mitbf{B}^{(\mathrm{e})}}

\newcommand{\uu}{\mitbf{v}}
\newcommand{\uuP}{\mitbf{v}^P}
\newcommand{\uuT}{\mitbf{v}^T}

\newcommand{\E}{\mitbf{E}}

\newcommand{\rr}{\mitbf{r}}

\newcommand{\BM}{\B_{0}}
\newcommand{\BG}{\B_{C}}
\newcommand{\BJ}{\B_{J}}

\newcommand{\Eimp}{\E^{\mathrm{imp}}}

\newcommand{\Bjm}[1]{B_{jm}^{(#1)}}

\newcommand{\Gejm}{G_{jm}^{(\mathrm{e})}}
\newcommand{\Gijm}{G_{jm}^{(\mathrm{i})}}

\newcommand{\Yjm}{Y_{jm}}
\newcommand{\Sjm}[1]{\mitbf{S}_{jm}^{(#1)}}

\newcommand{\pdt}[1]{\frac{\partial #1}{\partial t}}

\newcommand{\Br}{B_r}

\newcommand{\Dw}{D}
\newcommand{\Di}{H}

\newcommand{\be}{\begin{equation}}
\newcommand{\ee}{\end{equation}}

\journalname{JGR: Planets}

\begin{document}
\graphicspath{ {./} }

\let\WriteBookmarks\relax
\hyphenation{me-ri-dio-nal reccur-sion bench-marking dy-na-mi-cal de-pre-ssion con-fi-gu-ra-tion go-ver-ning using mea-nings mo-dels co-ming ele-va-tions se-lec-tive re-sul-ting di-ffe-ren-ces di-ffe-rent ma-nu-al ha-zard pro-pa-ga-tion}

% =========================================================
%                     TITLE AND AUTHORS
% =========================================================

%\title{Magnetic field induced by convective flow in %Ganymede’s subsurface ocean}
\title{Can the magnetic field induced by convective flow constrain the properties of Ganymede's subsurface ocean?}
%\shorttitle{Magnetic field induced by convective flow in Ganymede's subsurface ocean}
\authors{L. \v{S}achl\affil{1}, J. Kvorka\affil{1}, O. \v{C}adek\affil{1}, and J. Vel\'{i}msk\'{y}\affil{1}}

\affiliation{1}{Department of Geophysics, Faculty of Mathematics and Physics, Charles University}

\correspondingauthor{Libor \v{S}achl}{libor.sachl@mff.cuni.cz}

%%%\author[1]{Libor \v{S}achl}[orcid = 0000-0003-3281-3877]
%%%% Address/affiliation
%%%\affiliation[1]{organization = {Department of Geophysics, Faculty of Mathematics and Physics, %%%Charles University}, 
%%%                addressline  = {V Hole\v sovi\v ck\' ach 2}, 
%%%                city         = {Praha 8}, 
%%%                postcode     = {18000}, 
%%%                country      = {Czech Republic}}
%%%\cormark[1]  % Corresponding author indication
%%%\cortext[1]{Corresponding author}   % Corresponding author text
%%%\ead{libor.sachl@mff.cuni.cz}  % Email id of the first author
%%%\author[1]{Jakub Kvorka}[orcid = 0000-0002-9150-4524]
%%%\author[1]{Ond\v{r}ej \v{C}adek}[orcid = 0000-0001-8331-3093]
%%%\author[1]{Jakub Vel\'{i}msk\'{y}}[orcid = 0000-0002-1279-7112]

% =========================================================
%                      KEY POINTS
% =========================================================
%  List up to three key points (at least one is required)
%  Key Points summarize the main points and conclusions of the article
%  Each must be 140 characters or fewer with no special characters or punctuation, and must be complete sentences

\begin{keypoints}
\item Magnetic field generated by the flow (OIMF) in Ganymede's subsurface ocean could be strong enough to be measured by satellite missions.
\item Convective regime (mode) of the flow can be determined from the OIMF structure.
\item Ocean thickness can be derived from the latitudinal positions of the key features of the OIMF, provided the ice thickness is known.
%Ocean thickness can be derived from the ice thickness and the latitudinal positions of the key features of the ocean-induced magnetic field.
%\item Flow Mode IIa can be distinguished from Mode I and IIb using the respective OIMF patterns.
%OIMF pattern can distinguish flow Mode IIa from Mode I and IIb.
\end{keypoints}

% =========================================================
%                        ABSTRACT
% =========================================================

% Requirements (https://www.agu.org/publications/authors/journals/text-graphics-requirements):
% Be set as a single paragraph.
% Be less than 250 words for all journals except GRL, for which the limit is 150 words.
\begin{abstract}
This paper has two main objectives. First, we assess the strength and measurability of Ganymede's ocean-induced magnetic field (OIMF) generated by the flow in the subsurface ocean. Second, we inspect how the OIMF constrains the ocean flow and Ganymede's internal structure and suggest a suitable metric for this purpose. We calculate the OIMF by solving the electromagnetic induction (EMI) equation in the simplified form, where we neglect the ocean-induced field in the advection term and the motionless induction. We show that this approach is sufficiently accurate by comparison with the solution of the full EMI equation. We also demonstrate that the OIMF is predominantly a stationary signal generated by the interaction of the toroidal component of the flow with Ganymede's internal field. The contributions to the OIMF due to the poloidal component of the flow and Jupiter's magnetic field can be neglected as second-order effects. The stationary OIMF is a measurable quantity with a few nT amplitude on Ganymede's surface if the ocean is sufficiently thick ($\approx 400$~km) and conductive ($\approx 5$~S/m). 
In this case, the specific footprint of the measured OIMF could be used to identify the flow regime (mode) of the convecting ocean.
Additionally, the OIMF can be used to determine the ocean thickness since the OIMF pattern is shifted towards low latitudes if the ocean thickness is reduced.
%In this case, the measured OIMF could be used to identify the convecting ocean's flow regime (mode), as Mode IIa has a different magnetic footprint than the other two flow modes. 
%In Mode IIa, OIMF minima and maxima positions can also be used to determine the ocean thickness since the OIMF pattern is shifted towards low latitudes if the ocean thickness is reduced. 
%%%%Finally, we study the OIMF sensitivity to Ganymede's internal structure.  

%This paper is a continuation of the study by \citeA{Sachl2025}, who calculated the ocean-induced magnetic field (OIMF) generated by the flow in the subsurface ocean of Europa. The current study focuses on Jupiter's next icy moon, Ganymede.
\end{abstract}

\section*{Plain Language Summary}
In this paper, we study the magnetic field generated by the flow in Ganymede's saline and, thus, electrically conductive subsurface ocean. Our calculations suggest that, for a realistic Ganymede's interior, the constant-in-time component of the ocean-induced magnetic field (OIMF) could be strong enough to be measured by the upcoming Juice and Europa Clipper interplanetary missions. Of course, the parameters of the subsurface ocean, such as its flow regime, electrical conductivity, and thickness, significantly affect the strength and spatial pattern of the OIMF. We thus suggest a method to derive these parameters from the OIMF measurements. 
%%%The flow in Ganymede's subsurface ocean can also be partly constrained from the OIMF measurements since one of the three possible flow modes generates the OIMF signal with a significantly different pattern than the other two modes.

%%%\begin{keywords}
%%%Ganymede \sep topography \sep ocean dynamics
%%%\end{keywords}

% =========================================================
\section{Introduction} 
% ========================================================= 
% Please use ONLY \cite and \citeA for reference citations.
% use < > for prenotes and [ ] for postnotes
% DO NOT use other cite commands (e.g., \citet, \citep, \citeyear, \nocite, \citealp, etc.).

Ganymede is the only moon in the Solar System known to possess its own internal magnetic field \cite<GIF,>[]{kivelson1996,gurnett1996,kivelson1997a}. It has a strength of about 720 nT at the equator and 1440 nT at the poles \cite{Kivelson2002,Weber2022}, which is significantly more than the strength of the magnetic field of Jupiter (JMF) at the distance of Ganymede \cite<120 nT,>[]{kivelson1997a}. The magnetic field is likely to be generated by convection in Ganymede's metallic core, but the details of this process are still poorly understood \cite<e.g.,>[]{Hauck2006,Christensen2015a,Christensen2015b}. In addition to the intrinsic magnetic field, Ganymede has an induced magnetic field associated with the variations of the Jovian magnetic field along Ganymede's orbit. The induced magnetic field has a strength of about 60 nT at the equator and is considered evidence that Ganymede has a subsurface water ocean with a high electrical conductivity \cite{Kivelson2002}. 

Recently, \citeA{vance2021} have suggested that the convective flows of salt water in the presence of GIF may generate an additional magnetic field (often denoted by the acronym OIMF --- Ocean Induced Magnetic Field), and estimated that the strength of this field could range from 8 to 330~nT, depending on the flow speed and the composition and thickness of the ocean. An independent estimate of the OIMF by \citeA{kvorka2025}, based on a careful analysis of the flow speed in Ganymede's ocean, indicates that the magnetic signal generated by convection is an order of magnitude smaller than predicted by \citeA{vance2021} but within the sensitivity of the magnetic field measurements \cite{Kivelson2023,ellmaier2024}. In addition, \citeA{kvorka2025} show that the measurement of this signal can provide constraints on the geometry of ocean circulation. Owing to the presence of the intrinsic field, the magnetic field generated by convection in Ganymede's ocean is likely to be significantly stronger than the flow-induced magnetic field of Europa, where the flow interacts only with JMF. Based on the numerical simulations of convection in Europa's ocean and the solution of the electromagnetic
induction (EMI) equation, \citeA{Sachl2025} have estimated that the flow-induced magnetic field of Europa has a strength of $\lesssim\! 1$~nT, which is likely beyond the practical detectability by the Juice and Europa Clipper missions. 

While the research of the OIMF of Europa and Ganymede is still in its infancy, the magnetic signature of ocean circulation on Earth has been studied for more than 50 years \cite<e.g.,>[]{Sanford71,Larsen85,Vivier2004,Manoj2006}. The extraction of the tidally driven OIMF from the satellite data
\cite{Tyler2003,Grayver2019,Sabaka2020,Grayver2024} gave an impetus to the development of new methods and models \cite{Velimsky2018,Martinec2021}, leading to a better understanding of the distribution of electrical conductivity within the Earth \cite{Grayver2017,Sachl2022,Sachl2024}. The elusive component of OIMF, generated by the global ocean circulation driven by temperature and pressure gradients and interactions with the atmosphere, is yet to be detected in satellite magnetic data \cite{Finlay2024}. However, some of the numerical tools used to study the OIMF on Earth \cite<e.g.,>[]{Velimsky2018,Sachl2019,Velimsky2021b} can easily be modified and applied to study the subsurface oceans in icy moons \cite{Sachl2025}.  

A major challenge in studying the dynamics of subsurface oceans is that, unlike the oceans on Earth, the oceans in icy moons are not accessible to direct observation. The thickness of the ice layer above the ocean and the composition and depth of the ocean are poorly known \cite{vance2018}, and most of what we know about the flow in the subsurface oceans is based on the numerical modeling of thermal convection. The subsurface oceans differ from the oceans on Earth in three important respects. First, the oceans on icy moons are heated from below by the radiogenic heat coming from the silicate mantle or core. In contrast, the circulation of the oceans on Earth is primarily driven by solar energy. Second, the subsurface oceans lie beneath a layer of solid ice, unlike the Earth, where the ocean is in contact with the atmosphere. Finally, the thickness of the water layer on icy moons is likely to be of the order of tens to hundreds of km, and therefore much greater than on Earth. Numerical modeling of convection in the subsurface oceans has received growing attention in recent years. Particular focus has been given to Enceladus \cite<e.g.,>[]{kang2022a,kang2022b,bire2023}, Titan \cite<e.g.,>[]{amit2020,kvorka2022,terranova2023,kvorka2024} and Europa \cite<e.g.,>[]{soderlund2014,ashkenazy2021,ashkenazy2022,bire2023,lemasquerier2023,Sachl2025}. The dynamics of Ganymede's ocean has been briefly discussed in studies comparing the geometry of ocean circulation in different icy moons \cite<e.g.,>[]{soderlund2019} and has only recently been studied in detail \cite{kvorka2025}.     
 
In the present study, we use the models of Ganymede's ocean circulation developed by \citeA{kvorka2025} to investigate the OIMF in Ganymede's saline, electrically conductive ocean. The magnetic response of Ganymede is computed by solving the EMI equation, which includes the GIF and the changes in the JMF along Ganymede's orbit. To assess the strength and distribution of the induced magnetic field, we examine different ocean circulation models and vary the electrical conductivity and thickness of the ocean. This study aims to provide a preliminary estimate of the OIMF in Ganymede's ocean and to show under which conditions the magnetic measurement to be made by the Juice and Europa Clipper missions can provide information about the thickness of the ocean and the geometry of ocean circulation.

% =========================================================
%\section{Input information} 
% =========================================================

\section{Modelling of the ocean-induced magnetic field}
The OIMF in Ganymede is excited by the interaction of the thermally driven convection of
saline, electrically conductive water with the GIF maintained by dynamo action in Ganymede's
core, the JMF, and by the inductive response to the temporal variations of JMF.
In general, the EMI equation, derived from the quasi-static
approximation of Maxwell's equations and Ohm's law in moving
continuum, reads as,
\be
\label{EMI}
\mo\pdt{\Btot} + \curl\left(\frac{1}{\sigma}\curl\Btot\right) - \mo\curl\left(\uu\times\Btot\right) = \vz,
\ee
where $\Btot(\rr;t)$ is the total magnetic field,
$\sigma(\rr)$ is the electrical conductivity of Ganymede's
interior, $\mo$ is the magnetic permeability of free space, $\rr=(r,\vartheta,\varphi)$ is the position vector,
$\nabla$ is the 3-D gradient operator,
and $\uu(\rr;t)$ is the velocity field, non-zero only in the liquid regions of the moon.

Separation of the OIMF, further denoted as $\B$, from
the total magnetic field $\Btot$, requires a series
of linearizations, which are described in detail
in the Supplementary material with the corresponding boundary
conditions. In particular, we refer
to the \textit{Full solution}, which accounts for, and
eventually subtracts the magnetic field induced by temporal
variations of JMF along Ganymede's orbit. The
\textit{Simplifed solution} simply neglects it.
In both cases, we calculate the magnetic field using a
1-D, spherically symmetric variant of the time-domain,
spherical-harmonic finite-element
solver \cite{Sachl2019,Velimsky2021b,Sachl2025}.

% ---------------------------------------------------------
\section{Ocean circulation models}
% ---------------------------------------------------------

\begin{figure}[htb]
\centering
\includegraphics[width=0.65\textwidth]{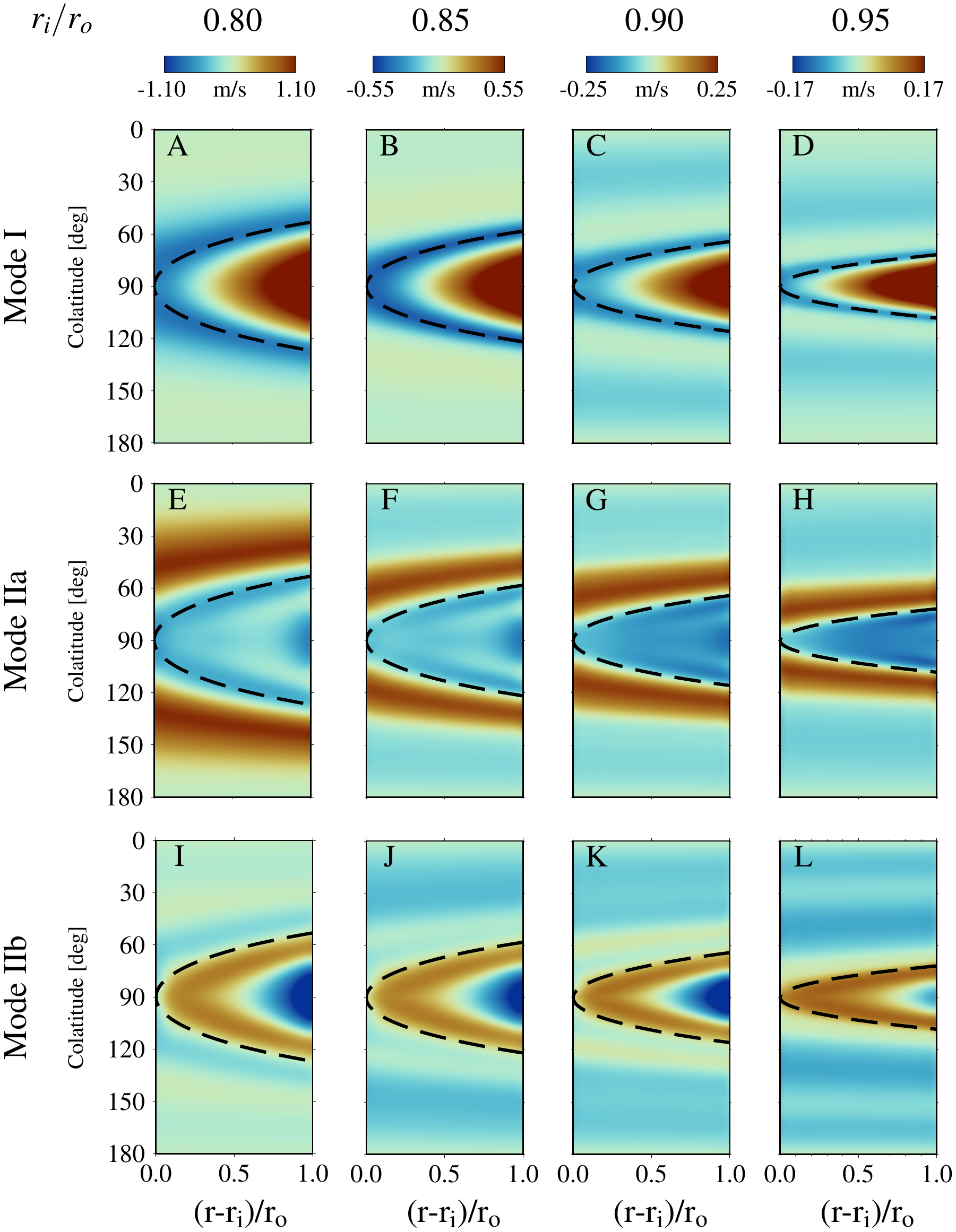}
\caption{Zonal flow in Ganymede's ocean for convection modes I, IIa and IIb \protect\cite{kvorka2025} and $r_i/r_o$ equal to $0.80$, $0.85$, $0.90$ and $0.95$. If the ice shell were $70$\,km thick, these radius ratios would correspond to ocean thicknesses of $513$, $385$, $256$, and $128$\,km, respectively. Ganymede's total heat production is assumed to be $1$\,TW. Black dashed lines show the position of the tangent cylinder (the cylinder of radius $r_i$ aligned with the rotation axis).
}
\label{models_flow}
\end{figure}

The magnetic field induced by the flow in Ganymede's ocean is calculated for ocean circulation models derived by \citeA{kvorka2025} from a large set of numerical convection simulations in three-dimensional spherical geometry. \citeA{kvorka2025} suggest that Ganymede's ocean is in the so-called transitional regime, meaning that the influence of rotation is relatively significant compared to buoyancy, but may no longer have a dominant role \cite<e.g.,>[]{gastine2016,cheng2018}. The flow pattern is characterized by the presence of strong zonal jets, alternating in direction and concentrated at low latitudes and, in some cases, extending throughout the entire volume of the ocean \cite{yadav2015,soderlund2019,bire2022,ashkenazy2021,Sachl2025}. Depending on the Ekman and Rayleigh numbers, the circulation can be dominated by either a prograde equatorial jet (Mode I, see Figure~\ref{models_flow}), or a retrograde equatorial jet and two prograde jets at mid-latitudes (Mode IIa), or a retrograde equatorial jet and a prograde jet at low latitudes (Mode IIb). Since the intersection between the tangent cylinder (dashed lines in 
 Figure~\ref{models_flow})  and the upper boundary of the ocean shifts towards the equator as the thickness of the ocean decreases, the meridional extent of the domain controlled by the zonal flow decreases with increasing $r_i/r_o$, where $r_i$ and $r_o$ are the radii of the inner and outer boundary, respectively.

The flow speed primarily depends on the heat flow from the deep interior, $Q$, and the thickness of the ocean, $D$ \cite<e.g.,>[]{bire2022, cabanes2024, Sachl2025, kvorka2025}. However, these parameters are poorly known, with estimates ranging from $200$ to $1600$\,GW \cite{choblet2017} and from $24$ to $518$\,km \cite{vance2018}, respectively. Considering the extreme cases and using the extrapolation method proposed by \citeA{kvorka2025}, we find that the root-mean-square speed can vary from $2$\,cm/s ($D=24$\,km, $Q=200$\,GW) to $78$\,cm/s ($D=518$\,km, $Q=1600$\,GW). Most of this variation is related to the changes in $D$: while increasing $Q$ from $200$\,GW to $1600$\,GW doubles the flow speed for a fixed value of $D$, reducing $D$ from $518$\,km to $24$\,km leads to an order of magnitude decrease in the flow speed for a constant $Q$. The zonal jet speed ranges from approximately $17$\,cm/s for $r_i/r_o = 0.95$ to $1.1$\,m/s for $r_i/r_o = 0.8$ (Figure ~\ref{models_flow}). These values are consistent with the estimate of \citeA{cabanes2024} based on the rapidly rotating limit (0.17--1.66~m/s). However, they are significantly lower than the speeds predicted for Ganymede by \citeA{vance2021}, who assumed that the flow speeds obtained from numerical simulations can be directly used to represent the flow in the oceans of icy satellites \cite<see also the discussion in>[]{Sachl2025}.

\clearpage
 
% ---------------------------------------------------------
\section{Ambient magnetic field}
% ---------------------------------------------------------

\begin{figure}[htb]
\centering
\includegraphics[width=\textwidth]{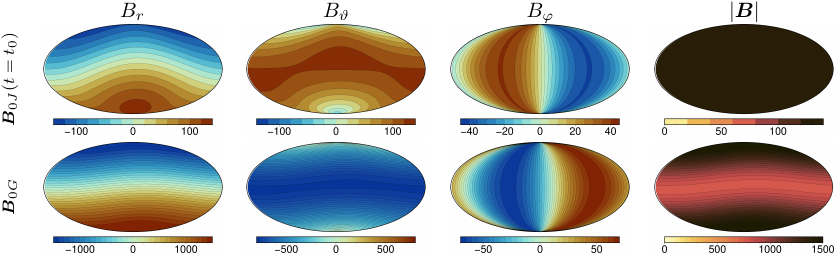}
\caption{Jupiter's magnetic field (upper panels) in initial time $t=t_0$ and Ganymede's internal magnetic field (lower panels) on Ganymede's upper ocean boundary.}
\label{B0_comparison}
\end{figure}

The ambient magnetic field is composed of the magnetic fields of Jupiter and Ganymede. JMF is prescribed by the spherical-harmonic (SH) model of \citeA{Connerney2022}. As in \citeA{Sachl2025}, we only use SH coefficients up to degree 13, although SH coefficients are provided up to SH degree 30. We also neglect spatial variations of JMF inside Ganymede since Ganymede's diameter is several orders of magnitude smaller than its distance from Jupiter. The model of \citeA{Connerney2022} does not provide secular variations of JMF.
%All SH coefficients are constant in time, the secular variations of JMF are not considered.
However, Ganymede experiences time-varying JMF with a synodic period of 10.53~h due to the misalignment between Jupiter's rotation and dipolar axes. JMF in Ganymede's spherical-coordinate frame in the initial time $t_0$, when Ganymede is in periapsis, and it is facing Jupiter's prime meridian, is depicted in the upper row of panels in Figure~\ref{B0_comparison}. 
%The JMF strength at the distance of Ganymede is approximately 120\,nT. 

The GIF generated in the moon's core is predominantly dipolar. The GIF dipole is tilted with respect to Ganymede's rotational axis by 176$^{\circ}$, meaning that it is directed against the magnetic moment of Jupiter \cite{Kivelson2002}. Its magnetic moment ($\approx 1.3\times 10^{20}$~A\,m$^2$) is orders of magnitude smaller than the Earth's magnetic moment but about three times larger than Mercury's. In our calculations, we use the model of \citeA{Weber2022} that provides degree one coefficients of Ganymede's dipole. The quadrupole coefficients are also available, but we do not use them since they are not well justified by the data, as the quadrupole moment is an order of magnitude weaker than the dipole term. Similar to the model of \citeA{Connerney2022}, no time evolution of the SH coefficients is considered in the GIF model of \citeA{Weber2022}. The GIF is depicted in the lower row of panels in Figure~\ref{B0_comparison}.

\section{Models of internal structure}
\label{sec:int_struct} 
% ---------------------------------------------------------

Ganymede's interior is described by a layered model. We consider five layers: iron core, silicate mantle, ice mantle, ocean, and ice. In all numerical experiments, the respective core and silicate mantle thicknesses are set to 700\,km and 1020\,km, which is in agreement with \citeA{Andersson2005}, \citeA{Bland2008}, and~\citeA{Kuskov2001}. The reference values of ice shell and ocean thicknesses are $\Di = 70$\,km and $\Dw = 361$\,km \cite{vance2018,kvorka2025}.

The electrical conductivity is constant in each layer. The reference conductivity values of core, silicate mantle, ice mantle, ocean, and ice shell are set to $\sigma_c = 10^3$\,S/m, $\sigma_m =10^{-2}$\,S/m, $\sigma_{mi} =10^{-5}$\,S/m, $\sigma_o =5$\,S/m and $\sigma_i =10^{-5}$\,S/m, respectively. I.e., we use the same reference conductivity values as \citeA{Sachl2025} for Europa. The exception is the ocean conductivity that we lowered from 10\,S/m to 5\,S/m since Europa's and Ganymede's oceans are expected to differ in composition and pressure--temperature conditions. In particular, Ganymede's ocean is expected to primarily be composed of MgSO$_4$, while Europa's ocean could contain MgSO$_4$ or NaCl (seawater). As the saturation-based upper conductivity limit for the NaCl composition is 18\,S/m but only 6\,S/m for the MgSO$_{4}$ composition \cite{Hand2007}, Ganymede's ocean is likely to be less conductive than Europa's. \citeA{vance2018} constructed two depth profiles of electrical conductivity in Ganymede assuming 10\,wt\% MgSO$_4$ composition. The conductivity values range from approximately 3 to 4.5\,S/m in the warm profile and from 2 to 2.6\,S/m in the cold profile. \citeA{Pan2020} conducted experimental electrical conductivity measurements and recalculated the conductivity profiles of \citeA{vance2018} using the same pressure--temperature conditions. The resulting conductivity profiles are more resistive than the conductivity profiles of \citeA{vance2018}. The conductivity values range from approximately 2.0 to 2.4\,S/m in the warm profile and from 1.25 to 1.6\,S/m in the cold profile. However, these values are still approximately one order of magnitude larger than the ocean conductivity predicted by \citeA{Jia2025}. Using magnetohydrodynamic simulations, they refined the strength of the induced magnetic field by subtracting the contributions of plasma and ionospheric currents from the magnetic measurements of Galileo and Juno. Their results suggest that the efficiency of the magnetic induction should be reduced from 0.84 to 0.72. Consequently, the ocean conductivity may drop as low as approximately 0.08~S/m for a 200~km thick ocean and a 150~km thick ice shell. Note that this result agrees with the estimate of \citeA{Saur2015}, who studied the orientation of Ganymede's auroral oval. It is beyond the scope of this paper to examine the order-of-magnitude difference in the conductivity estimates of \citeA{vance2018, Pan2020} and \citeA{Saur2015, Jia2025}. Still, the reader should keep in mind that the chosen value of the reference ocean conductivity probably represents the corresponding upper limit.

Ganymede's internal structure is uncertain. The thickness and conductivity of each layer may vary within acceptable limits. Thus, we test the sensitivity of OIMF to different conductivity and thickness values of Ganymede's internal layers in Section~\ref{sec:GOIMF_structure}. For example, ice is expected to be resistive, ice conductivity above $10^{-2}$\,S/m would require a partially melted ice \cite{Zimmer2000, Keller1966}. Accordingly, we consider ice conductivity in the range of $10^{-6}$\,S/m \cite{Marusiak2021} and $10^{-3}$\,S/m. In contrast, the ocean is expected to be conductive, so we consider the ocean conductivity in the range of 0.5\,S/m and 5\,S/m. 
%but the true ocean conductivity is likely to be lower. 
We also consider different ice shell and ocean thicknesses according to \citeA{kvorka2025}.

%%%These ranges are taken into consideration, and the sensitivity of OIMF to the different conductivity values is tested in Section~\ref{sec:GOIMF_structure}.

% =========================================================
\section{Results}   % Ganymede's OIMF
% =========================================================
% Use lowercase letters (a, b,c...) to label parts of the figure; do not use Arabic or Roman numerals.

We calculate the OIMF in several configurations listed in Table~\ref{GOIMF_configurations}. The individual configurations are specified by the complexity of the solution represented by the applied modeling approach (see Supplementary Material) and the choice of input fields - the ambient magnetic field and the velocity field.   

The OIMFs presented in Sections~\ref{sec:GOIMF_physics} and~\ref{sec:GOIMF_jets_cells} are calculated using Mode~IIb, which is the most probable flow mode in Ganymede's ocean \cite{kvorka2025}. In Section~\ref{sec:GOIMF_modes}, we present the OIMFs generated by all flow modes.
\begin{table}[htb]
\centering
\begin{tabular}{cccc}
\hline
Configuration  &  Method       &  Ambient magnetic field  &  Flow    \\ \hline
1              &  Simplified   &  $\BJ$                   &  $\uu$   \\
2              &  Simplified   &  $\BG$                   &  $\uu$   \\
3              &  Full         &  $\BG+\BJ(t)$            &  $\uu$   \\
4              &  Simplified   &  $\BG$                   &  $\uuT$  \\
5              &  Simplified   &  $\BG$                   &  $\uuP$  \\
\end{tabular}
\caption{Summary of configurations used to calculate OIMFs in Figures~\ref{GOIMF_physics} and~\ref{GOIMF_flow_comps}.}
\label{GOIMF_configurations}
\end{table}

% ---------------------------------------------------------
\subsection{Impact of the ambient magnetic field and the solution complexity}     \label{sec:GOIMF_physics} 
% ---------------------------------------------------------

\begin{figure}[htb]
\centering
\includegraphics[width=\textwidth]{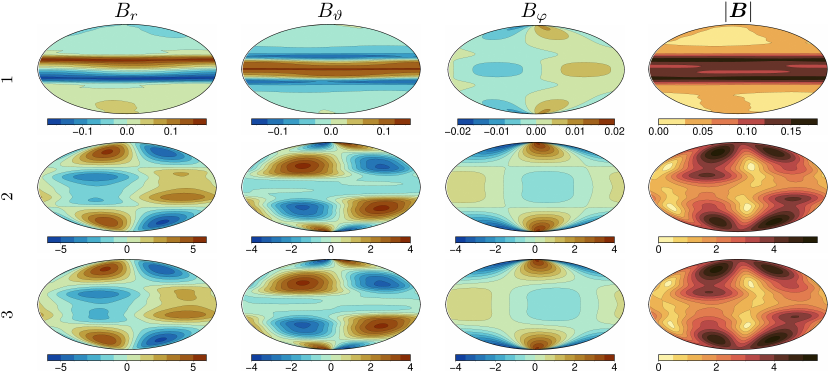}
\caption{OIMFs [nT] in Configurations~1-3 calculated using the reference thickness model. The OIMF in Configuration~1 is calculated in time $t=t_{spin}+4$~h.}
\label{GOIMF_physics}
\end{figure}

We start with the Configuration~1 used in \citeA{Sachl2025}. The ambient magnetic field is given by the JMF, and the OIMF is obtained by solving the EMI equation using the Simplified approach (see Supplementary Material). The OIMF calculated in time $t=t_{spin}+4$~h, where $t_{spin}=1$\,year is a sufficiently long spin-up time, is shown in the upper row of panels in Figure~\ref{GOIMF_physics}. It is a very weak signal that does not reach 0.5~nT in amplitude. The result corresponds to the calculations of \citeA{Sachl2025} if we consider that Ganymede is larger than Europa, the ocean is approximately 2.4 times thicker, but the JMF amplitude is approximately 5 times smaller because Ganymede is more distant from Jupiter than Europa, and the ocean flow is only slightly faster (approximately 20\%).         

However, unlike Europa, Ganymede has its internal magnetic field $\BG$. According to Figure~\ref{B0_comparison}, this field can not be neglected since it is stronger than the time-varying JMF and may serve as the constant ambient magnetic field. The Simplified OIMF solution calculated in Configuration~2 is shown in the middle row of panels in Figure~\ref{GOIMF_physics}. It is a much stronger signal than the OIMF in Configuration~1, with an amplitude over 5\,nT.

Finally, we inspect whether the Simplified OIMF solution is accurate enough. For comparison, we calculate the Full OIMF solution. The OIMF in Configuration~3 is shown in the lower row of panels in Figure~\ref{GOIMF_physics}. If we compare OIMFs in Configurations~2 and~3, the differences are relatively subtle. The OIMF in Configuration~3 is locally distorted (in particular the zonal component) and slightly shifted in the zonal direction with respect to the OIMF in Configuration~2. However, the large-scale patterns of both OIMFs are the same. 

To sum up, the key characteristics of the OIMF can be calculated using the Simplified method with the ambient field given by Ganymede's field. This approach is also numerically cheap since the solution is stationary; there is no evolution in time. We thus use this approach throughout the rest of the paper. However, remember that the Simplified and Full OIMF solutions may differ locally.
% The JMF can be safely neglected as it generates only a small perturbation of the OIMF.      

% ---------------------------------------------------------
\subsection{OIMF generated by zonal jets and convection cells}   \label{sec:GOIMF_jets_cells} 
% ---------------------------------------------------------

\citeA{Sachl2025} have shown that the OIMF generated by the convection cells on Europa is equally strong as the OIMF generated by the zonal jets. This holds, although the zonal velocities in Europa's ocean reach approximately one order of magnitude larger values than the meridional/radial velocities in the convection cells. On Ganymede, the convection cells are almost missing in the flow Mode~I but are reasonably well developed and comparable to Europa's case in the flow Modes IIa and IIb (see Figure~\ref{models_flow}). It motivates us to study the OIMF sensitivity to different components of the flow field. 

\begin{figure}[htb]
\centering
\includegraphics[width=\textwidth]{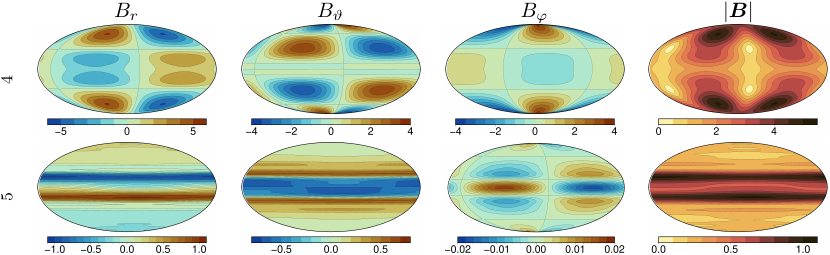}
\caption{OIMFs [nT] in Configurations~4 and~5 calculated using the reference thickness model.}
\label{GOIMF_flow_comps}
\end{figure}

In particular, we split Configuration~3 into Configurations~4 and~5, in which the velocity field is represented only by zonal jets with the toroidal velocity $\uuT$ and convection cells with the poloidal velocity $\uuP$, respectively. The corresponding OIMFs are shown in Figure~\ref{GOIMF_flow_comps}. The OIMF patterns are significantly different, which is a direct consequence of the very different flow patterns. However, the key point and the main difference with respect to Europa's OIMF is that the zonal jets dominate not only the flow field but also the OIMF signal. The OIMF generated by the zonal jets is approximately five times larger than the OIMF generated by the convection cells. The amplitude difference is even more pronounced in the zonal component. 

The different behavior of the OIMF on Europa and Ganymede can be attributed to the distinct character of the ambient magnetic field. In Europa's case, the ambient field is constant in the moon's interior. Thus, it is straightforward to decompose the ambient field into vertical and horizontal components, which display different behavior. In Ganymede's case, the ambient field is a time-invariant dipolar field generated by its internal dynamo, which implies that its decomposition to vertical and horizontal components can, in principle, be done, but it does not bring new insight into the problem.

% ---------------------------------------------------------
\subsection{OIMF generated by different flow modes}     \label{sec:GOIMF_modes} 
% ---------------------------------------------------------

\begin{figure}[htb]
\centering
\includegraphics[width=\textwidth]{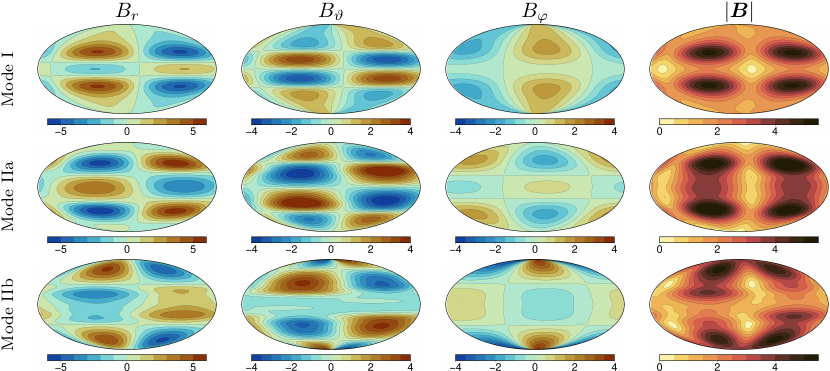}
\caption{OIMFs [nT] generated by Modes I, IIa, and IIb in the reference thickness model.}
\label{GOIMF_flow_modes}
\end{figure}

\citeA{kvorka2025} identified three distinct flow modes in Ganymede's subsurface ocean. The corresponding OIMFs are depicted in Figure~\ref{GOIMF_flow_modes}. Regarding the OIMF pattern, Mode~I and Mode~IIb generate qualitatively similar OIMFs, while Mode~IIa generates the OIMF with opposite signs of the key features. The Ganymede's OIMF could thus be used to identify the ocean flow in Mode~IIa easily. To explain this behaviour, recall that we have shown in Sec.~\ref{sec:GOIMF_jets_cells} that the zonal jets dominate the OIMF. Outside of the tangent cylinder (the cylinder of radius $r_i$ that is aligned with the rotation axis), 
the zonal jets are dominated by the retrograde flows in Mode~IIa, which is different in Modes~I and~IIb.
%the zonal jets are dominated by the prograde flows in Modes~I and~IIb, while there are retrograde flows in Mode~IIa. 
%The Ganymede's OIMF could thus be used to identify the ocean flow in Mode~IIa easily. 
If the flow is in Mode~I or Mode~IIb, we may still be able to distinguish between them using the positions of the local OIMF extrema. For further details, see Section\,\ref{sec:imprint} and the discussion related to Figure~\ref{B_avz}. Regarding the OIMF amplitude, the strongest signal is generated by Mode~IIa, and the signals from the other two modes are slightly weaker. Specifically, the largest $\left| \Br \right|$ values calculated using Mode~IIa and Modes~I and~IIb are 5.8~nT and 5.2~nT, respectively. The OIMF may thus be a measurable signal by the Juice space probe. 

\begin{figure}[htb]
\centering
\includegraphics[width=0.8\textwidth]{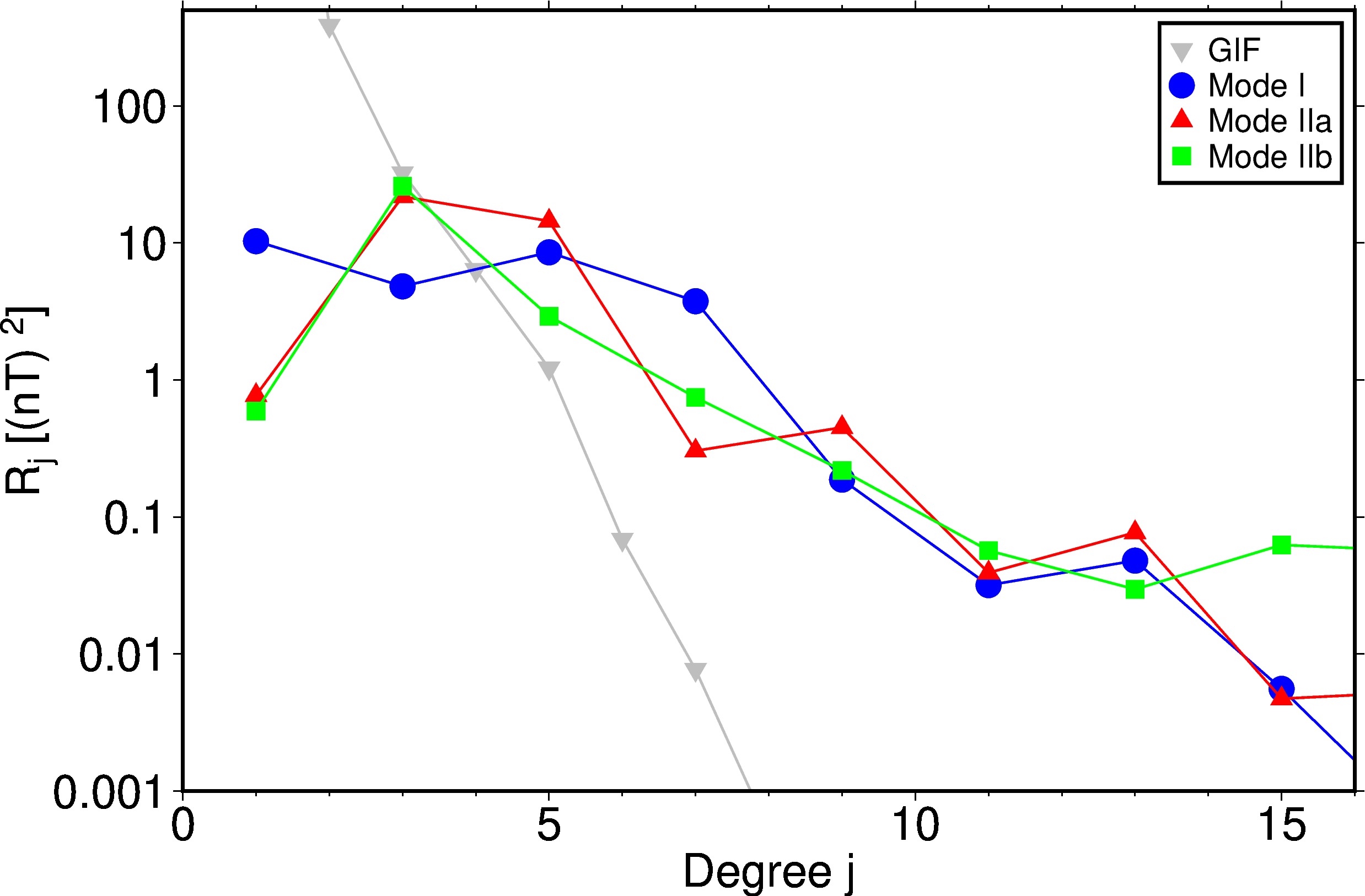}
\caption{Power spectra of the GIF and three OIMFs. We do not plot even degrees of the OIMF power spectra because there is no power on them.}
\label{power_spectra}
\end{figure}

On the other hand, as discussed in Section~\ref{sec:GOIMF_physics}, the leading component of the OIMF is represented by a stationary field, and a complication may arise when separating this signal from the GIF, which varies only very slowly over time. However, the spatial patterns of both fields are distinct (compare Figures~\ref{B0_comparison} and~\ref{GOIMF_flow_modes}), and this difference can be used to separate them. Regarding power spectra, we are interested in the spectral region where the OIMF is more energetic than the GIF. The respective Lowes--Mauersberger power spectra $R_j$ are shown in Figure~\ref{power_spectra}. The GIF power spectrum has been derived using the normalized power spectrum for case 14.4 in \citeA{Christensen2015a} that we recalculated on Ganymede's surface and $R_1 = 1.016 \times 10^6$\,(nT)$^2$ \cite{Kivelson2002}. As expected, the predominantly dipolar GIF dominates the SH degree $j=1$, but the OIMF has more power on degrees $j \geq 5$.  In fact, the true decrease of the GIF spectrum may be steeper, as the power spectrum in (Christensen 2015a) was calculated using the defective code. The GIF's $R_2/R_1$ ratio decreased when calculated using the corrected code (Christensen 2015b).   
%The dominant SH degree is $j=3$ in Modes~IIa and~IIb. In Mode~I the largest power is on $j=1$ but $j=5$ is comparably strong.

% ---------------------------------------------------------
\subsection{OIMF sensitivity to the Ganymede's interior structure}     \label{sec:GOIMF_structure} 
% ---------------------------------------------------------

\begin{figure}[htb]
\centering
\includegraphics[width=\textwidth]{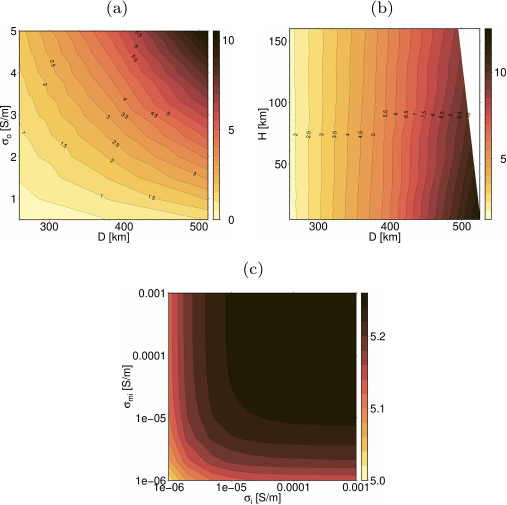}
\caption{OIMF [nT] sensitivity to (a) ocean thickness and conductivity, (b) ocean and ice thicknesses, and (c) conductivities of ice shell and ice mantle.}
%%%\caption{OIMF sensitivity [nT] to the conductivities of ice shell and ice mantle (left panel) and to the ocean and ice thicknesses (right panel).}
\label{GOIMF_sensitivity}
\end{figure}

In this section, we perform a sensitivity study to assess the uncertainties in the OIMF due to the loosely constrained structural parameters. Since the ocean flow drives an OIMF, the parameters of the ocean layer are of paramount importance. Figure~\ref{GOIMF_sensitivity}a shows the sensitivity of the OIMF strength to ocean thickness and conductivity. The strength of the OIMF increases as the ocean becomes thicker and/or more conductive, consistent with \citeA{Sachl2025}. In particular, the OIMF strength is almost linearly dependent on ocean conductivity. The dependence of OIMF strength on ocean thickness is more complex. The OIMF strength increases more rapidly with the ocean thickness if the ocean is thinner. 
For example, if $\sigma_o=5$\,S/m, the OIMF increases from 1.7\,S/m to 5.2\,S/m and 10.4\,S/m if $\Di=70$\,km and $\Dw = 256$\,km, 385\,km and 513\,km, respectively.
%For example, if $\sigma_o=5$\,S/m, the OIMF increases from 1.7\,S/m to 7.2\,S/m and 8.4\,S/m if $\Dw = 250$\,km, 380\,km and 510\,km, respectively.
%For example, if $\sigma_o=5$\,S/m, the OIMF increases from 1.72\,S/m to 7.7\,S/m and 9.5\,S/m if $\Dw = 250$\,km, 380\,km and 510\,km, respectively.
%with higher ocean conductivity and    

Like ocean thickness, ice thickness is another key parameter that is not fully constrained. Figure\,\ref{GOIMF_sensitivity}b shows the sensitivity of the OIMF strength to ocean and ice thickness. In most cases, thicker ice results in a weaker OIMF because the source is shifted away from the receiver \cite{Sachl2025}. However, it is somewhat surprising that ice thickness affects the OIMF strength more when the ocean is thicker. Moreover, if the ocean is thicker, the OIMF strength increases more rapidly with the ocean thickness if the ice is thin.

Finally, Figure\,\ref{GOIMF_sensitivity}c shows the sensitivity of the OIMF strength to the conductivities of the ice shell and ice mantle. The OIMF strength does not change much in the inspected range of conductivity values. The largest OIMF values are obtained for the most conductive ice shell and ice mantle. Note that if we consider only the GIF, the ice conductivity does not appear in the EMI equation in the approximate approach since the EMI equation reduces to the Laplace equation in all layers except for the ocean layer. The ice conductivity affects the OIMF only through the boundary condition on the ocean-ice interface,
%%%\be
%%%\mitbf{n} \times \left[ \frac{\nabla \times \B}{\sigma} \right]_{-}^{+} = \mo \mitbf{n} \times (\uu \times \BM),
%%%\ee
which expresses the requirement of continuity of the lateral component of the ocean-induced electric field.

% ---------------------------------------------------------
\subsection{Imprint of the ocean thickness}     \label{sec:imprint} 
% ---------------------------------------------------------

Assume for a moment that the ice shell thickness is known from the other geophysical measurements (e.g., tidal deformation). Figure\,\ref{GOIMF_sensitivity}a shows that the strength of the OIMF (its maximum value) may then constrain the ocean thickness or the ocean conductivity if the complementary quantity is known. But can we constrain one of these quantities without knowing the complementary one? As Figure\,\ref{models_flow} shows, the thinning of the ocean layer, equivalent to decreasing the radius ratio $r_i/r_o$, shifts the boundaries of the tangent cylinder towards lower latitudes. Consequently, the flow outside the tangent cylinder becomes more confined to lower latitudes if we decrease the ocean thickness. It stands to reason that the OIMF could be affected similarly.

\begin{figure}[htb]
\centering
\includegraphics[width=\textwidth]{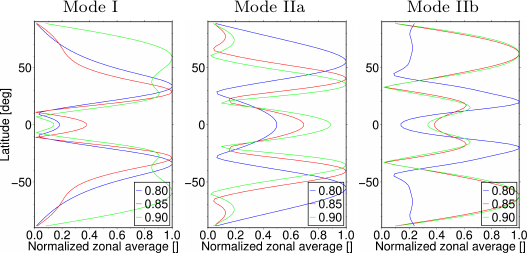}
\caption{Zonally averaged $\left| \B\right|$ field [nT] generated by Modes~I, IIa and IIb using radius ratios $r_i/r_o$ of 0.80, 0.85 and 0.90.}
\label{B_avz}
\end{figure}

\begin{figure}[htb]
\centering
\includegraphics[width=0.5\textwidth]{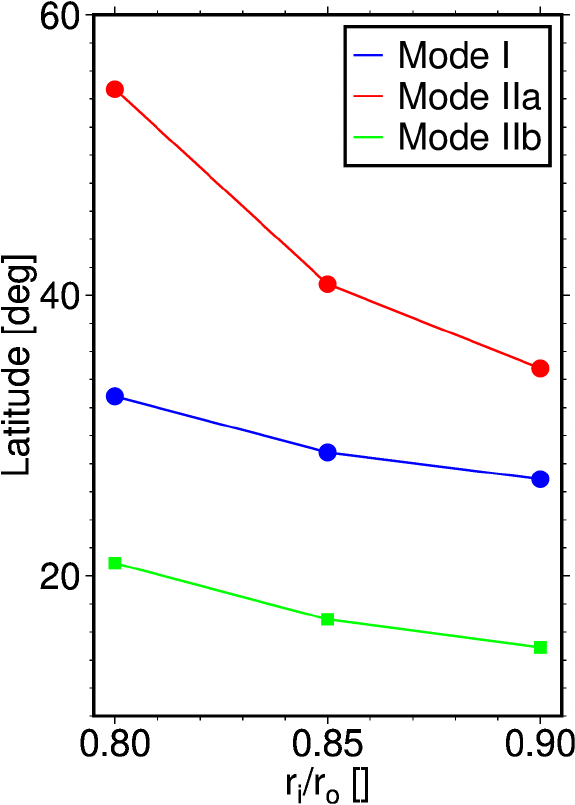}
\caption{Positions of maxima in the zonally averaged radial component of the OIMF in absolute value [nT] as a function of the radius ratio $r_i/r_o$ calculated using the flow Modes\,I, IIa, and IIb.}
\label{thickness_meter}
\end{figure}

In order to make the phenomenon apparent, we zonally average the OIMF in absolute value and normalize it. Figure\,\ref{B_avz} depicts the resulting meridional profiles calculated for the radial component, flow Modes\,I, IIa, and IIb, and $r_i/r_o$ ratios of 0.80, 0.85, and 0.90. Let us start with Mode~IIa, in which the shape of the curve depends only weakly on the $r_i/r_o$ ratio. The meridional profile is gradually ``compressed'' towards the equator with increasing $r_i/r_o$ as expected from the inspection of the velocity field. If the upcoming satellite missions could measure the compression level, it could be used to constrain the ocean thickness via $r_i/r_o$. In particular, we propose locating the latitudinal position of the peak closest to the equator. As Figure\,\ref{thickness_meter} shows, $r_i/r_o$ ratios of 0.80, 0.85, and 0.90 could be distinguished from each other. The respective latitudinal positions of the peaks are 54.7$^{\circ}$, 40.8$^{\circ}$ and 34.8$^{\circ}$. In Mode\,I, the latitudinal positions of the peaks are 32.8$^{\circ}$, 28.8$^{\circ}$ and 26.9$^{\circ}$, respectively. Thus, it might be difficult to distinguish between 0.85 and 0.90 ratios. On the other hand, as Figure\,\ref{GOIMF_sensitivity}a suggests, if $r_i/r_o=0.90$, the OIMF is most likely too weak to be measured by the satellite missions anyway. In Mode\,IIb, the latitudinal positions of the peaks are 20.9$^{\circ}$, 16.9$^{\circ}$ and 14.9$^{\circ}$. I.e., similar to Mode\,I, the separability of 0.85 and 0.90 ratios could be complicated.
%Figure\,\ref{thickness_meter} shows that the latitudinal position of the maximum of the zonally averaged $\left| \B\right|$ field shifts towards the equator (low latitudes) with the increasing $r_i/r_o$ ratio.
%--------
%In Modes\,I and\,IIb, we apply the same methodology to the radial component. In Mode\,I, the latitudinal positions of the peaks for 0.80, 0.85, and 0.90 ratios are 32.8$^{\circ}$, 28.8$^{\circ}$ and 26.9$^{\circ}$, respectively. Thus, it might be difficult to distinguish between 0.85 and 0.90 ratios. On the other hand, as Figure\,\ref{GOIMF_sensitivity}a suggests, if $r_i/r_o=0.90$, the OIMF is most likely too weak to be measured by the satellite missions anyway. In Mode\,IIb, the latitudinal positions of the peaks for 0.80, 0.85, and 0.90 ratios are 20.9$^{\circ}$, 16.9$^{\circ}$ and 14.9$^{\circ}$, respectively. I.e., the separability of 0.85 and 0.90 ratios could be complicated, similar to Mode\,I.

We discussed in Section~\ref{sec:GOIMF_modes} that Mode\,IIa can be easily identified using its characteristic OIMF pattern. We also mentioned that Modes\,I and\,IIb can be separated using the specific OIMF features. Figure\,\ref{B_avz} is suitable for that purpose: Mode\,I has a local maximum at the equator, which is followed by a local minimum approximately at $10^{\circ}$ latitude (see the left panel in Figure\,\ref{B_avz}). In contrast, Mode\,IIb has a local minimum at the equator, followed by a local maximum (we used it to determine the $r_i/r_o$ ratio). To summarize, each possible flow mode can be identified using the OIMF.

% =========================================================
\section{Conclusions}
% =========================================================

We have calculated the magnetic field induced by convective flow (OIMF) in Ganymede's subsurface ocean. Our results suggest that the OIMF provides a valuable tool to constrain the properties of the subsurface ocean. We can identify the convective regime (flow mode) using the OIMF pattern and the local extrema in the zonally averaged OIMF in absolute value. We also proposed a method to determine the ocean thickness from the OIMF, provided the ice thickness is known. On the other hand, we anticipate specific difficulties when extracting the OIMF from the satellite measurements. The most severe issue is the strength of the OIMF. For realistic ocean and ice thicknesses of 385\,km and 70\,km, respectively, the OIMF reaches values over 5\,nT at Ganymede's surface if the ocean conductivity is 5\,S/m. However, if ocean conductivity is one order of magnitude smaller, as suggested by some recent studies, the OIMF will not be detectable as its strength depends almost linearly on the ocean conductivity. Another issue is that the OIMF signal and Ganymede's core field are predominantly constant in time; thus, separating them can be challenging. However, the advantage is that the spatial patterns of  Ganymede's core field and the OIMF differ. Specifically, the OIMF calculated for the reference internal-structure model is stronger than the GIF on the spherical-harmonic degrees five and higher.

% =========================================================
\appendix
%%%\renewcommand\theequation{\Alph{section}.\arabic{equation}}
%%%\renewcommand{\thetable}{\Alph{section}.\arabic{table}}
%%%\renewcommand{\thefigure}{\Alph{section}.\arabic{figure}}

% =========================================================

\clearpage

\section*{Open Research Section}
% =========================================================
%This section MUST contain a statement that describes where the data supporting the conclusions can be obtained. Data cannot be listed as "Available from authors" or stored solely in supporting information. Citations to archived data should be included in your reference list. Wiley will publish it as a separate section on the paper's page. Examples and complete information are here:
%https://www.agu.org/Publish with AGU/Publish/Author Resources/Data for Authors

Modeling results can be downloaded from 
%\url{https://zenodo.org/records/14725726}.

% =========================================================
\section*{As Applicable – Inclusion in Global Research Statement}
% =========================================================
The Authorship: Inclusion in Global Research policy aims to promote greater equity and transparency in research collaborations. AGU Publications encourage research collaborations between regions, countries, and communities and expect authors to include their local collaborators as co-authors when they meet the AGU Publications authorship criteria (described here: https://www.agu.org/publications/authors/policies\#authorship). Those who do not meet the criteria should be included in the Acknowledgement section. We encourage researchers to consider recommendations from The TRUST CODE - A Global Code of Conduct for Equitable Research Partnerships (https://www.globalcodeofconduct.org/) when conducting and reporting their research, as applicable, and encourage authors to include a disclosure statement pertaining to the ethical and scientific considerations of their research collaborations in an ``Inclusion in Global Research Statement'' as a standalone section in the manuscript following the Conclusions section. This can include disclosure of permits, authorizations, permissions and/or any formal agreements with local communities or other authorities, additional acknowledgements of local help received, and/or description of end-users of the research. You can learn more about the policy in this editorial. Example statements can be found in the following published papers: 
Holt et al. (https://agupubs.onlinelibrary.wiley.com/doi/full/10.1029/2022JG007188), 
Sánchez-Gutiérrez et al. (https://agupubs.onlinelibrary.wiley.com/doi/abs/10.1029/2023JG007554), 
Tully et al. (https://agupubs.onlinelibrary.wiley.com/doi/epdf/10.1029/2022JG007128) 
Please note that these statements are titled as ``Global Research Collaboration Statements'' from a previous pilot requirement in JGR Biogeosciences. The pilot has ended and statements should now be titled "Inclusion in Global Research Statement".

% =========================================================
\acknowledgments
% =========================================================
This research was supported by the Czech Science Foundation, project No. 25-16801S, the European Space Agency (ESA), contract No. 4000141625/23/NL/MGu/my, and the Ministry of Education, Youth, and Sports of the Czech Republic through the e-INFRA CZ (ID:90254). J.K. acknowledges the support from the Charles University project SVV 260709. O.\v{C}. is a member of the Nečas Center for Mathematical Modeling. J.V. acknowledges the funding by ESA Contract No. 4000143627/24/I-EB Swarm for Ocean Dynamics, part of the ESA Solid Earth Magnetic Science Cluster, EO for Society programme.
%We also thank two anonymous reviewers for their helpful comments. 

% =========================================================
%                       REFERENCES
% =========================================================
%\clearpage
%\bibliographystyle{agu}
\bibliography{bibliography}

\end{document}

\graphicspath{ {./Images/} }

\let\WriteBookmarks\relax
\hyphenation{me-ri-dio-nal reccur-sion bench-marking dy-na-mi-cal de-pre-ssion con-fi-gu-ra-tion go-ver-ning using mea-nings mo-dels co-ming ele-va-tions se-lec-tive re-sul-ting di-ffe-ren-ces di-ffe-rent ma-nu-al ha-zard pro-pa-ga-tion}

% =========================================================
%                     TITLE AND AUTHORS
% =========================================================

\title{Magnetic field induced by convective flow in Ganymede’s subsurface ocean\\
Supplementary material}
%\shorttitle{Magnetic field induced by convective flow in Ganymede’s subsurface ocean}
\authors{L. \v{S}achl\affil{1}, J. Kvorka\affil{1}, O. \v{C}adek\affil{1}, and J. Vel\'{i}msk\'{y}\affil{1}}

\affiliation{1}{Department of Geophysics, Faculty of Mathematics and Physics, Charles University}

\correspondingauthor{Libor \v{S}achl}{libor.sachl@mff.cuni.cz}

%%%\author[1]{Libor \v{S}achl}[orcid = 0000-0003-3281-3877]
%%%% Address/affiliation
%%%\affiliation[1]{organization = {Department of Geophysics, Faculty of Mathematics and Physics, %%%Charles University}, 
%%%                addressline  = {V Hole\v sovi\v ck\' ach 2}, 
%%%                city         = {Praha 8}, 
%%%                postcode     = {18000}, 
%%%                country      = {Czech Republic}}
%%%\cormark[1]  % Corresponding author indication
%%%\cortext[1]{Corresponding author}   % Corresponding author text
%%%\ead{libor.sachl@mff.cuni.cz}  % Email id of the first author
%%%\author[1]{Jakub Kvorka}[orcid = 0000-0002-9150-4524]
%%%\author[1]{Ond\v{r}ej \v{C}adek}[orcid = 0000-0001-8331-3093]
%%%\author[1]{Jakub Vel\'{i}msk\'{y}}[orcid = 0000-0002-1279-7112]

% =========================================================
%                      KEY POINTS
% =========================================================
%  List up to three key points (at least one is required)
%  Key Points summarize the main points and conclusions of the article
%  Each must be 140 characters or fewer with no special characters or punctuation and must be complete sentences

% =========================================================
\section{Linearization of the EMI equation}
\label{sec:modelling_EMI}
% =========================================================
The magnetic field in the Ganymede's interior is a sum of 
contributions of different origins and characteristics.
The core field is generated by the self-consistent dynamo action
in the moon's liquid metallic core. The ocean-induced magnetic field is
produced by the kinematic dynamo in the liquid sub-surface ocean,
where the magnetic feedback on the ocean flow is neglected.
The temporal variations of the external field, stemming from the
interactions between the Jupiter's field of internal origin,
its magnetosphere and the magnetosphere of Ganymede, penetrate
from the surface to the moon's interior. All components are
mutually linked by inductive and galvanic interactions
in the strongly conductive metallic and water layers,
and weakly conductive silicate and ice layers.

The total magnetic field in the Ganymede's interior $\Btot(\rr;t)$ is described by the
EMI equation, derived in turn from the quasi-static approximation of the 
Maxwell equations, and the Ohm's law in moving continuum,
\be
\label{EMI}
\mo\pdt{\Btot} + \curl\left(\frac{1}{\sigma}\curl\Btot\right) - \mo\curl\left(\uu\times\Btot\right) = \vz,
\ee
where $\sigma(\rr)$ is the electrical conductivity of Ganymede's
interior, $\mo$ is the magnetic permeability of free space, $\rr=(r,\vartheta,\varphi)$ is the position vector,
$\nabla$ is the 3-D gradient operator,
and $\uu(\rr;t)$ is the velocity field, non-zero only in the liquid regions of the moon.
Above the moon surface, $r\ge a$, the magnetic field
is described by a scalar magnetic potential $\bar{U}(\rr;t)$, satisfying the
Laplace equation and therefore represented by a spherical-harmonic series,
separating the magnetic fields generated respectively by sources above and below the surface,
\be
\bar{U}(\rr;t) = \bar{U}^{\mathrm{(e)}}(\rr;t) + \bar{U}^{\mathrm{(i)}}(\rr;t)= a \sum\limits_{j=1}^\infty\sum\limits_{m=-j}^j
\left[\Gejmtot(t)\left(\frac{r}{a}\right)^j + \Gijmtot(t)\left(\frac{a}{r}\right)^{j+1}\right]\Yjm(\vartheta,\varphi),
\ee
where $\Gejmtot(t)$ and $\Gijmtot(t)$ denote respectively the spherical harmonic
coefficients of external and internal field, and $\Yjm(\vartheta,\varphi)$
are real, fully normalized spherical harmonic functions.
The continuity of magnetic field across the Ganymede's surface,
\be
\left.\Btot = -\grad\bar{U}\right|_{r=a},
\ee
implies two boundary conditions for the EMI equation solution inside the moon.
The toroidal magnetic field disappears, 
\be
\BtotT(r=a,\vartheta,\varphi;t) = \vz,
\ee
and the external field of Jovian and mangetospheric origin,
represented by the $\Gejmtot(t)$ spherical harmonic coefficients, is imposed,
\be
\Btotext(r=a,\vartheta,\varphi;t)= \BE(\vartheta,\varphi;t).
\ee

Given the estimates of the diffusion and advection times of the Ganymede's dynamo of
the order of $10^4$ and $10^2$ years, respectively \cite{Christensen2015a}, for the
purpose of ocean and induction studies which act on shorter time scales,
the core field $\Bc$ can be considered as a stationary potential
field outside the core domain, $r\ge c$,
\be
\pdt{\Bc}= 0, \quad \Bc=-\grad U_\mathrm{C}.
\ee
Influence of the external fields, and ocean-induced fields on the core field is neglected.
This leads directly to the first linearization of the EMI equation, describing the
sum of ocean-induced and externally induced fields $\Boj=\Btot-\Bc$, with 
a-priori imposed forcing term on the right-hand side,
\begin{eqnarray}
\label{EMIall}
\mo\pdt{\Boj} + \curl\left(\frac{1}{\sigma}\curl\Boj\right) - \mo\curl\left(\uu\times\Boj\right) & = & \mo\curl\left(\uu\times\Bc\right),\\
\label{BCall1}
\BojT(r=a,\vartheta,\varphi;t) & = & \vz,\\
\label{BCall2}
\Bojext(r=a,\vartheta,\varphi;t) & = & \BE(\vartheta,\varphi;t).
\end{eqnarray}

The second linearization separates the ``motionless'' induction process
from the motionally-induced magnetic field.
Let $\Bjs$ be the solution of the externally driven EMI equation
in the moon at rest,
\begin{eqnarray}
\label{EMIstat}
\mo\pdt{\Bjs} + \curl\left(\frac{1}{\sigma}\curl\Bjs\right) & = & \vz,\\
\label{BCstat1}
\BjsT(r=a,\vartheta,\varphi;t) & = & \vz,\\
\label{BCstat2}
\Bjsext(r=a,\vartheta,\varphi;t) & = & \BE(\vartheta,\varphi;t).
\end{eqnarray}

Subtracting respectively Equations (\ref{EMIstat}--\ref{BCstat2}) from
Equations (\ref{EMIall}--\ref{BCall2}), yields for the motionally-induced
field $\Bo=\Boj-\Bjs$,
\begin{framed}
\begin{eqnarray}
\label{EMIocean}
\mo\pdt{\Bo} + \curl\left(\frac{1}{\sigma}\curl\Bo\right) - \mo\curl\left(\uu\times\Bo\right) & = &
\mo\curl\left[\uu\times\left(\Bc+\Bjs\right)\right],\\
\label{BCocean1}
\BoT(r=a,\vartheta,\varphi;t) & = & \vz,\\
\label{BCocean2}
\Boext(r=a,\vartheta,\varphi;t) & = & \vz.
\end{eqnarray}
We refer to the solution of problem (\ref{EMIocean}--\ref{BCocean2}) as the \textit{Full solution}
in the main text.
\end{framed}

Further simplifications are possible. If the ocean-induced field is sufficiently small,
$|\Bo|\ll|\Bc|$, it can be neglected in the advection term in Equation (\ref{EMIocean}).
A-posteriori analysis of our results indeed shows that this effect will likely be below
10 \%.
Moreover, for slowly varying external fields, induction can be neglected, and the Jovian field
can be approximated by a potential field inside Ganymede,
\be
\pdt{\BE} \approx  \vz \Rightarrow \pdt{\Bjs} = \vz, \quad \Bjs  = -\grad \widehat{U}=\BJ.
\ee

Then the EMI equation for OIMF then reduces to
\begin{framed}
\begin{eqnarray}
\label{EMIoceanAppr}
\mo\pdt{\Bo} + \curl\left(\frac{1}{\sigma}\curl\Bo\right) & = &
\mo\curl\left(\uu\times\BM\right),\\
\label{BCoceanAppr1}
\BoT(r=a,\vartheta,\varphi;t) & = & \vz,\\
\label{BCoceanAppr2}
\Boext(r=a,\vartheta,\varphi;t) & = & \vz,
\end{eqnarray}
where the ambient magnetic field $\BM$ now stands for the Ganymede's core field $\Bc$,
the Jovian field $\BJ$, or their sum.
For time-invariant flow  $\uu\ne\uu(t)$ and ambient
field $\BM\ne\BM(t)$, the problem (\ref{EMIoceanAppr}--\ref{BCoceanAppr2}) converges to a stationary
solution as $t\rightarrow\infty$, which can be
obtained directly by setting $\frac{\mu_0}{\Delta t} = 0$
in the discrete time integration scheme.
The \textit{Simplified solution} (\ref{EMIoceanAppr}--\ref{BCoceanAppr2}) is assessed in Section~6.1.
\end{framed}

% =========================================================
%                       REFERENCES
% =========================================================
%\clearpage
%\bibliographystyle{agu}
\bibliography{bibliography}